\documentclass[twocolumn,noshowpacs,preprintnumbers,amsmath,amssymb]{revtex4}
\usepackage{graphicx}
\usepackage{dcolumn}
\usepackage{bm}

\begin{document}

\title{Surface superconductivity as the primary cause of broadening of superconducting transition in Nb-films}

\author{A. Zeinali}
\author{V. M. Krasnov}
\email{Vladimir.Krasnov@fysik.su.se}

\affiliation{$^1$Department of Physics, Stockholm University,
AlbaNova University Center, SE-10691 Stockholm, Sweden }


\begin{abstract}

We study the origin of broadening of superconducting transition in
sputtered Nb films. From simultaneous tunneling and transport
measurements we conclude that the upper critical field $H_{c2}$
always corresponds to the bottom of transition $R\sim 0$, while
the top $R\sim R_n$ occurs close to the critical field for
destruction of surface superconductivity $H_{c3}\simeq 1.7
H_{c2}$. The two-dimensional nature of superconductivity at
$H>H_{c2}$ is confirmed by cusp-like angular dependence of
magnetoresistance. Our data indicates that surface
superconductivity is remarkably robust even in disordered
polycrystalline films and, surprisingly, even in perpendicular
magnetic fields.


\end{abstract}

\maketitle

\section{Introduction}

Superconductivity occurs as a result of the second-order phase
transition, accompanied by a sudden appearance of the
superconducting order parameter below the critical temperature
$T_c$ and the upper critical field $H_{c2}$ \cite{SaintJames}.
This should lead to an abrupt vanishing of resistance. However, in
reality resistive transitions are always broadened, especially in
magnetic field. This is usually ascribed to flux-flow phenomenon
caused by motion of Abrikosov vortices \cite{BardeenStephen_1965}.
Broadening can also be caused by spatial inhomogeneity (e.g.
variation of $T_c$), or superconducting fluctuations
\cite{Varlamov_book,Finkelstein_2012,Glatz_2014}, particularly for
high temperature superconductors. Finally, surface
superconductivity (SSC) may survive up to a significantly higher
field $H_{c3}\simeq 1.69 H_{c2}$ than bulk superconductivity
\cite{SaintJames}, which can also smear the superconducting
transition. Although SSC is quite profound in polished clean
superconductors
\cite{Bellau_1966,Park_2004,Casalbuoni_2005,Das_2008}, it is
usually ignored for disordered, polycrysstaline films because SSC
is considered to be very sensitive to the quality of the surface
(e.g., surface passivation \cite{Casalbuoni_2005} and order
parameter suppression \cite{SaintJames}), surface roughness
\cite{Bellau_1966,Casalbuoni_2005} and surface scattering
\cite{Agterberg_1996}.

The broadening is detrimental for superconducting devices such as
transition-edge sensors \cite{TES1,TES2} and resonators
\cite{Casalbuoni_2005}. Presence of several mechanisms makes the
interpretation of broadening ambiguous. The lack of understanding
does not allow confident extraction of fundamental parameters of
superconductors, such as $H_{c2}$, because it is unclear which
point at the transition curve corresponds to $H=H_{c2}$. Arbitrary
criteria, such as 10, 50 or 90 $\%$ of the normal state resistance
$R_n$, are commonly applied which apparently does not work for
high-$T_c$ superconductors with very broad transitions
\cite{Zavaritsky}. Therefore, clarification of the mechanism of
broadening is important both for fundamental and applied research
on superconductors.

In this work we study the origin of broadening of superconducting
transitions in sputtered Nb films. We perform simultaneous
tunneling spectroscopy and transport measurements, which allow
unambiguous ascription of $H_{c2}$ to the bottom of resistive
transition $R(H_{c2})/R_n \sim 0$. The top of transition
corresponds to $\sim 1.7$ times higher fields, which is close to
the third critical field $H_{c3}$ for destruction of surface
superconductivity. The two-dimensional (2D) nature of SSC at
$H_{c2}<H<H_{c3}$ is confirmed by observation of cusp-like angular
dependence of magnetoresistance. Thus we conclude that surface
superconductivity, rather than flux-flow, inhomogeneity or
fluctuations, is the primary cause of broadening of
superconducting transitions in magnetic field. Our data indicates
that surface superconductivity is remarkably robust even in
disordered polycrystalline films and, surprisingly, even in
perpendicular magnetic fields.

\begin{figure*}[t]
    \centering
    \includegraphics[width=0.9\textwidth]{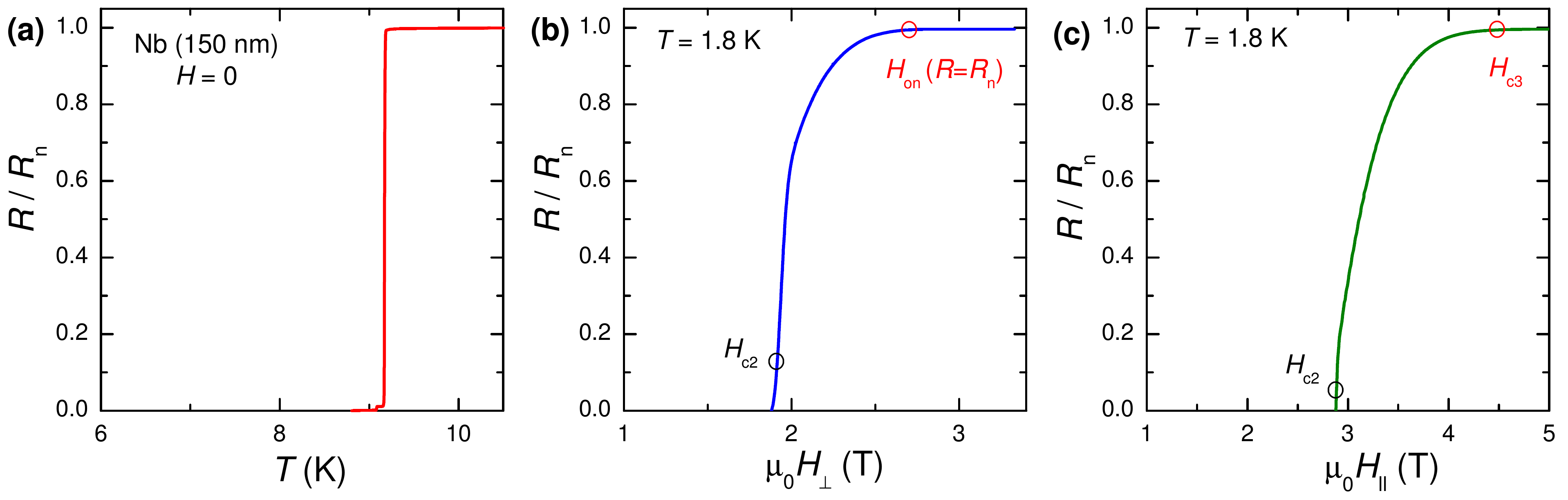}
    \caption{(Color online). Resistive transitions of a 150 nm thick Nb film. (a) Temperature dependence of
    the resistance in zero magnetic field. (b) and (c) Field dependencies of
    resistances at $T=1.8$ K in fields (b) perpendicular and (c)
    parallel to the film. Black and red circles mark the upper
    critical field $H_{c2}$ and the field of the onset of resistive
    transition, which coincides with the critical field of surface
    superconductivity $H_{on} \sim H_{c3}$.
    }
    \label{fig:fig1}
\end{figure*}

\section{Experimental}

The studied sample contains several Nb/Al-AlOx/Nb tunnel junctions
with sputtered Nb electrodes of thicknesses $d=150$ and 50 nm.
Junction characteristics in perpendicular fields were reported in
Ref. \cite{MR}. Due to different thicknesses, electrodes have
slightly different $T_c$ of 9.2 and 8.8 K. Parameters extracted
from tunneling characteristics are determined by the thinner
electrode, while transport measurements are made at the thicker
electrode. This explains a minor difference in $H_{c2}$ values
obtained by those techniques. Measurements are performed in a
gas-flow $^4$He cryostat with a superconducting solenoid.
Samples are mounted on a rotatable holder with the alignment
accuracy $\sim 0.02^{\circ}$. Details of the setup can be found
elsewhere \cite{MR}.

\section{Results}

In Figures \ref{fig:fig1} (a-c) we show superconducting
transitions of a 150 nm thick Nb film: (a) $R(T)$ in zero field
and $R(H)$ at $T=1.8$ K for field perpendicular (b) and parallel
(c) to the film. It is seen that at zero field the $R(T)$
transition is very sharp and does not show any extended
fluctuation region or inhomogeneity. However, $R(H)$ transitions
are quite broad. Interestingly, $R(H)$ is broader when the field
is parallel to the film. This confronts interpretation of
broadening in terms of vortex motion because the driving Lorentz
force is most effective in perpendicular and vanishes in parallel
field. Therefore, this broadening is not consistent with either
flux-flow, inhomogeneity or fluctuation mechanisms.

\begin{figure*}[th]
    \centering
    \includegraphics[width=1.0\textwidth]{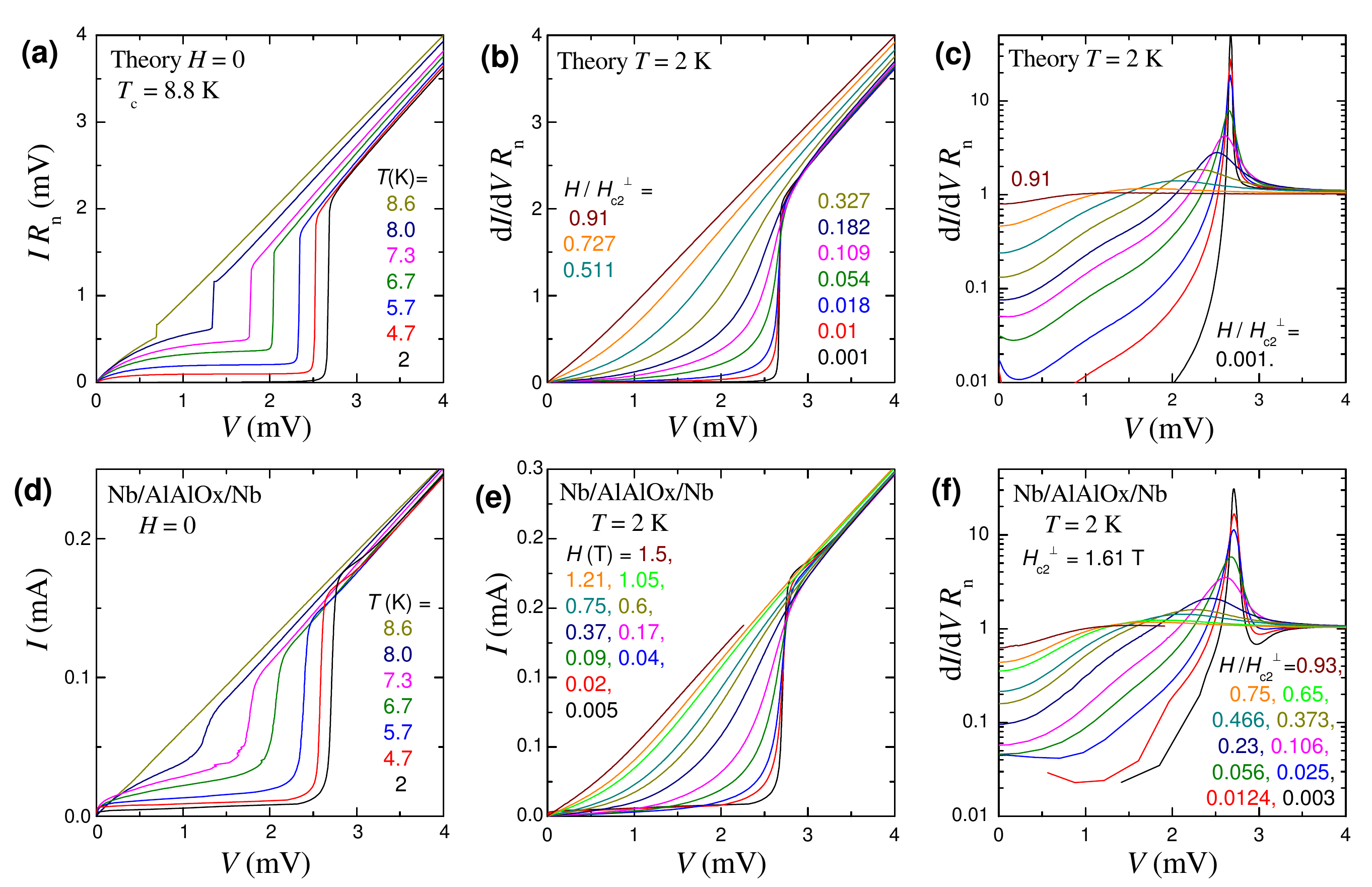}
    \caption{(Color online). Comparison of theoretically calculated (a-c) and experimentally measured (d-f)
    characteristics of Nb/AlAlOx/Nb tunnel junctions. (a) and (d) Temperature dependence of $I$-$V$ characteristics at zero field.
    (b) and (e) Field dependence of $I$-$V$ characteristics for field perpendicular to the junction/films at $T\simeq 2$ K.
    (c) and (f) The corresponding differential conductances for curves from panels (b) and (e).
    The field scale in (f) is normalized by the upper critical
    field $H_{c2}^{\perp}=1.61$ T, which is obtained as a single
    scaling factor for all the curves at different fields. Data
    from Ref. \cite{MR}.
    }
    \label{fig:fig2}
\end{figure*}

\subsection{Determination of $H_{c2}$ from tunneling spectroscopy}

In order to analyze surface superconductivity scenario, first of
all, it is necessary to determine bulk $H_{c2}$. For this we
perform magneto-tunneling spectroscopy. Figure \ref{fig:fig2}
represents a comparison of theoretically calculated (a-c) and
experimentally measured tunneling characteristics of our
Nb/AlAlOx/Nb junction (data from Ref. \cite{MR}). Details of
calculations are described in Ref. \cite{MR}). Panels (a) and (d)
show temperature dependencies of $I$-$V$ characteristics at zero
field. Panels (b) and (e) show field dependence of $I$-$V$
characteristics for field perpendicular to the junction/films at
$T\simeq 2$ K. Panels (c) and (f) show the corresponding
differential conductances for $I$-$V$ curves from panels (b) and
(e). There is a good quantitative agreement between theoretical
and experimental characteristics. The main spectroscopic features
are the sharp sum-gap peak at $V_p=2\Delta/e$, where $\Delta$ is
the superconducting gap and the suppressed quasiparticle current
and conductance below the sum-gap voltage. With increasing field
the peak is decreasing in height and is moving to lower voltages.
Simultaneously the sub-gap conductance is increased. All this is
due to suppression of the superconducting gap by magnetic field.
The extent of suppression depends solely on $H/H_{c2}$. Above
$H_{c2}$ the superconducting gap vanishes and the $I$-$V$ becomes
linear (Ohmic). Thus, the ratio $H/H_{c2}$ uniquely determines the
shape of tunneling characteristics in magnetic field. Therefore,
as discussed in Ref. \cite{MR}, the ratio $H/H_{c2}$ can be
uniquely determined from analysis of the shape of tunneling
characteristics. In Fig. \ref{fig:fig2} (f) the field is
normalized by thus obtained $H_{c2}^{\perp}=1.61$ T. We emphasize
that this value is obtained as a single fitting parameter for the
whole set of d$I/$d$V(V)$ characteristics at different $H$. This
removes the uncertainty in determination of $H_{c2}$.

\begin{figure*}[t]
    \centering
    \includegraphics[width=0.9\textwidth]{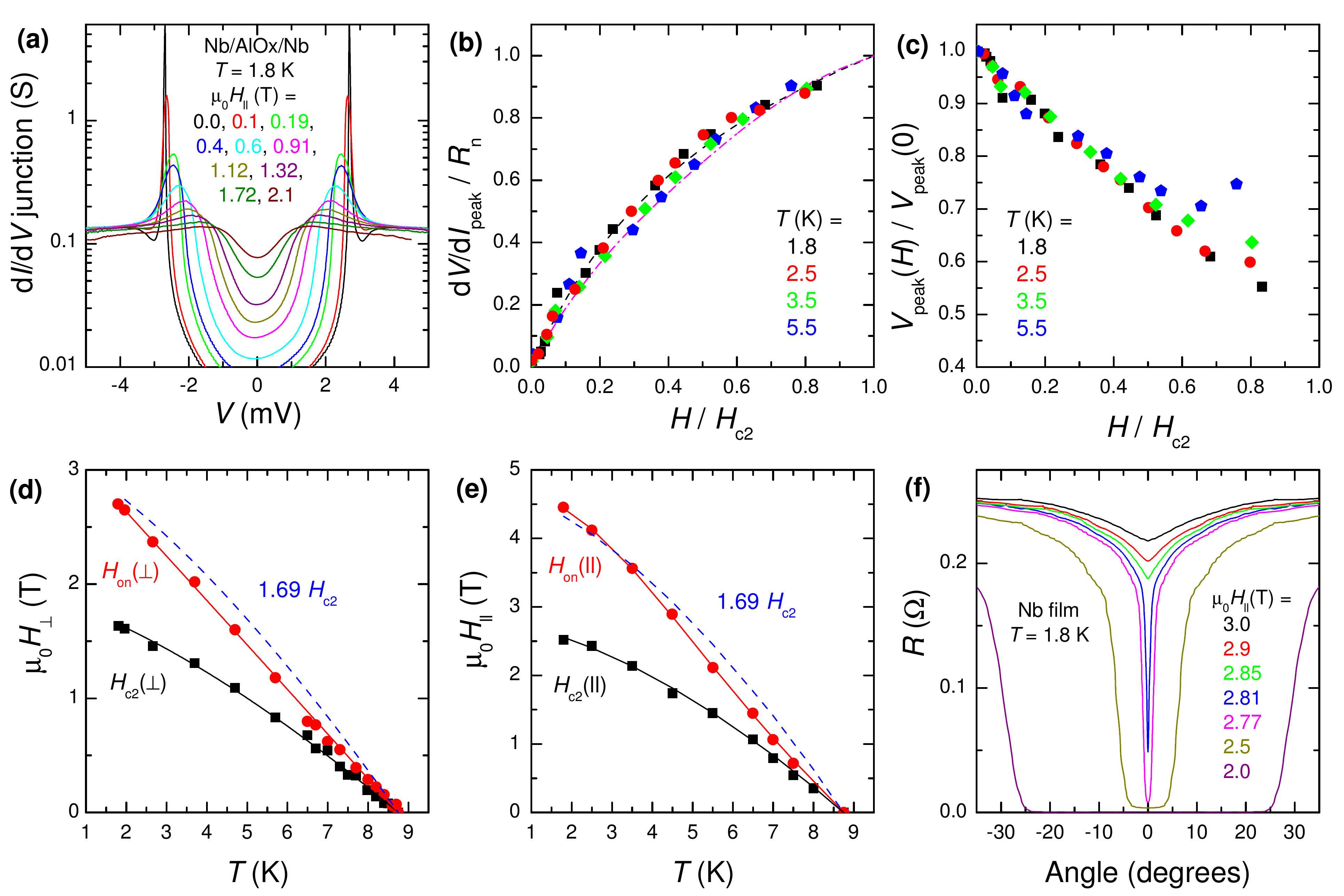}
    \caption{(Color online). (a) Differential conductances of a Nb/AlOx/Nb junction
    (in a semi-logarithmic scale) at parallel to the films magnetic fields and $T=1.8$ K.
(b) and (c) Scaling of the sum-gap peak resistance (b) and voltage
(c) as a function of $H/H_{c2}(T)$ at different temperatures and
parallel fields. Dashed and dashed-dotted lines in (b) represent
theoretical curves at $T=1.96$ K and 4.7 K, respectively. (d) and
(e) Black squares represent upper critical
    fields perpendicular (d) and parallel (e) to the films,
    obtained from the scaling of magneto-tunnelling
    characteristics. Dashed lines represent the expected third critical field for surface
    superconductivity $H_{c3}=1.69 H_{c2}$. Red circles mark the top onset of the resistive
    transition.(f) Angular dependence of resistance of Nb electrodes at fields slightly
    below and above $\mu_0 H_{c2}(\parallel)=2.52$ T. A cusp-like feature at $H>H_{c2}$ indicates occurrence of the
    two-dimensional surface superconductivity.
    }
    \label{fig:fig3}
\end{figure*}

Fig. \ref{fig:fig3} (a) shows a set of tunneling d$I$/d$V(V)$
characteristics of a Nb/AlOx/Nb junction at $T=1.8$ K in fields
parallel to Nb films. From comparison of Figs. \ref{fig:fig2} and
\ref{fig:fig3} (a), it can be seen that the influence of magnetic
field is qualitative similar both for parallel and perpendicular
field orientations. In Ref. \cite{MR} it was shown that the peak
height and the peak voltage exhibit universal almost
$T$-independent quasi-linear scaling as a function of
$H/H_{c2}(T)$. Fig. \ref{fig:fig3} (b) and (c) demonstrate such a
scaling at different temperatures for field parallel to Nb films.
Dashed and dashed-dotted lines in Fig. \ref{fig:fig3} (b)
represent theoretical results from Ref. \cite{MR} for $T=1.96$ and
4.7 K, correspondingly. The overall quality of scaling is quite
good, which allows confident extraction of $H_{c2}(T)$.
Thus obtained $H_{c2}$ is unambiguous because it is deduced as a
single fitting parameter for the whole set of d$I/$d$V(V)$
characteristics at fixed $T$ for different $H$.

Squares in Figs. \ref{fig:fig3} (d) and (e) represent obtained
$H_{c2}(T)$ dependencies for perpendicular and parallel field
orientations, respectively. Using the relation
$H_{c2}^{\perp}=\Phi_0/2\pi\xi^2$ we calculate the coherence
length $\xi_0\simeq 14$ nm. This small value indicates that the
film is in the dirty limit with a very short scattering length due
to a disordered film structure with nm-scale crystallites. Thus,
the studied Nb film $d=150$ nm is an order of magnitude thicker
than $\xi_0$.
This leads to an important for a further discussion conclusion
that our films are bulk three-dimensional (3D) superconductors
practically in the whole temperature range $T<T_c$. Red circles in
Figs. \ref{fig:fig1} (b) and (c) represent top onsets
$R(H_{on})\simeq R_n$ of resistive transitions.
Dashed blue lines correspond to $H_{c3} =1.69 H_{c2}$ expected for
surface superconductivity, which is close to the onset field.
Remarkably this is true even for the perpendicular field
orientation when SSC in the uniform case is not expected
\cite{SaintJames}.

\begin{figure*}[t]
    \centering
    \includegraphics[width=0.9\textwidth]{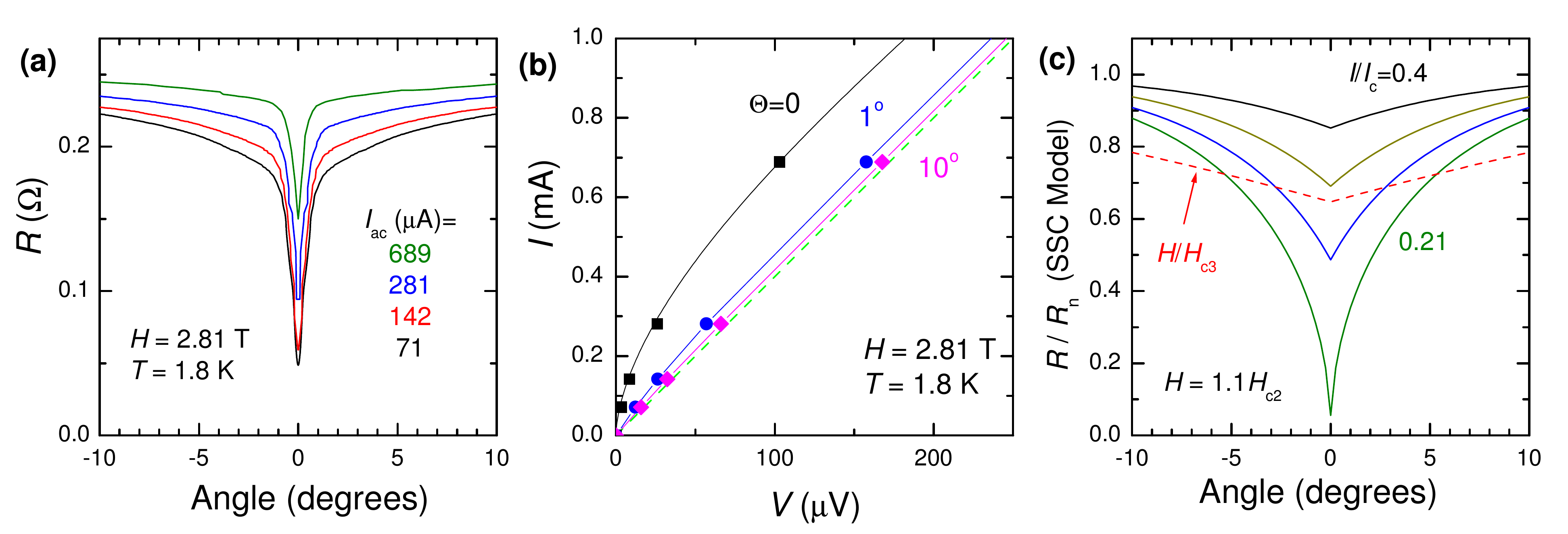}
    \caption{(Color online). (a) Bias dependence
    of $R(\Theta)$ at $T=1.8$ K and $H=2.81$ T. (b) Deduced
    current-voltage characteristics at different angles from the data in panel
    (a). Dashed line represents the normal state $I$-$V$.
    (c) Simulated angular dependence of $R(\Theta)$ in the
    surface superconductivity model, taking into account
    non-linearity of $I$-$V$ characteristics in the vicinity of
    critical current. Solid lines represent calculated of $R(\Theta)$
    at different bias. Dashed curve represents the standard
    flux-flow dependence $H/H_{c3}$.
    }
    \label{fig:fig4}
\end{figure*}

\subsection{Angular dependence of magnetoresistance}

The 2D nature of SSC should be reflected in a cusp-like angular
dependence of $H_{c3}$, given by the equation
\cite{Yamafuji_1966}:
\begin{equation}\label{Hc3_Q}
\left[\frac{H_{c3}(\Theta)}{H_{c3}^\parallel}\cos\Theta\right]^2A(\Theta)
+ \left|\frac{H_{c3}(\Theta)}{H_{c3}^\perp}\sin\Theta\right| =1,
\end{equation}
where $A(\Theta)=1+(1-\sin \Theta)\tan\Theta$. It is only slightly
different from Tinkham's 2D result with $A(\Theta)=1$.

Fig. \ref{fig:fig3} (f) shows angular-dependencies of
magnetoresistance at $T=1.8$ K and at different fields below and
above $\mu_0 H_{c2}^{\parallel} = 2.52$ T. Zero angle $\Theta=0$
corresponds to field parallel to the film. It is seen that at 2.5
T very slightly below $H_{c2}$, $R(\Theta)$ is flat at $\Theta=0$,
which is characteristic for 3D bulk Nb. However, at
$H>H_{c2}^{\parallel}$ angular dependence acquires a 2D cusp.
Since the film thickness is much larger than $\xi$, the observed
2D behavior at low $T$ may originate solely from SSC.

\subsection{Non-linear bias dependence}

The sheet surface critical current (in A/cm) is
\cite{Abrikosov_1965} :
\begin{equation}\label{JcH}
I_c \simeq \frac{5}{2\sqrt{3\pi}}\frac{H_c}{\kappa}
\left(1-\frac{H}{H_{c3}}\right)^{3/2}.
\end{equation}
Here $H_c$ is the thermodynamic critical field (in Oe) and
$\kappa$ is the Ginzburg-Landau parameter. Typically such $I_c$ is
in the range from few to few tens of A/m. For our films with
$\kappa \gg 1$ and the width of few microns the $I_c$ is in the
$\mu$A range, comparable to the probe current. Therefore, the
results do depend on the bias, as illustrated in Fig.
\ref{fig:fig4} (a). This is due to strong non-linearity of
current-voltage characteristics at $I\sim I_c$, as demonstrated in
Fig. \ref{fig:fig4} (b). In order to demonstrate how such the
non-linearity affects experimental characteristics we consider a
standard shape of $I$-$V$:
\begin{equation}\label{IV}
V=R_n\sqrt{I^2-I_c^2}.
\end{equation}
Together with Eq. (\ref{JcH}) it yields
\begin{equation}\label{R_Q}
1-\left[\frac{R(\Theta)}{R_n}\right]^2=\frac{I_c^2}{I^2}\left[1-\frac{H}{H_{c3}(\Theta)}\right]^{\nu}.
\end{equation}
The exponent $\nu$ depends on the shape of the $I$-$V$ and the
quality of the surface \cite{Bellau_1966}. For the case of Eq.
(\ref{IV}) it is $\nu=3$.

\begin{figure*}[t]
    \centering
    \includegraphics[width=0.6\textwidth]{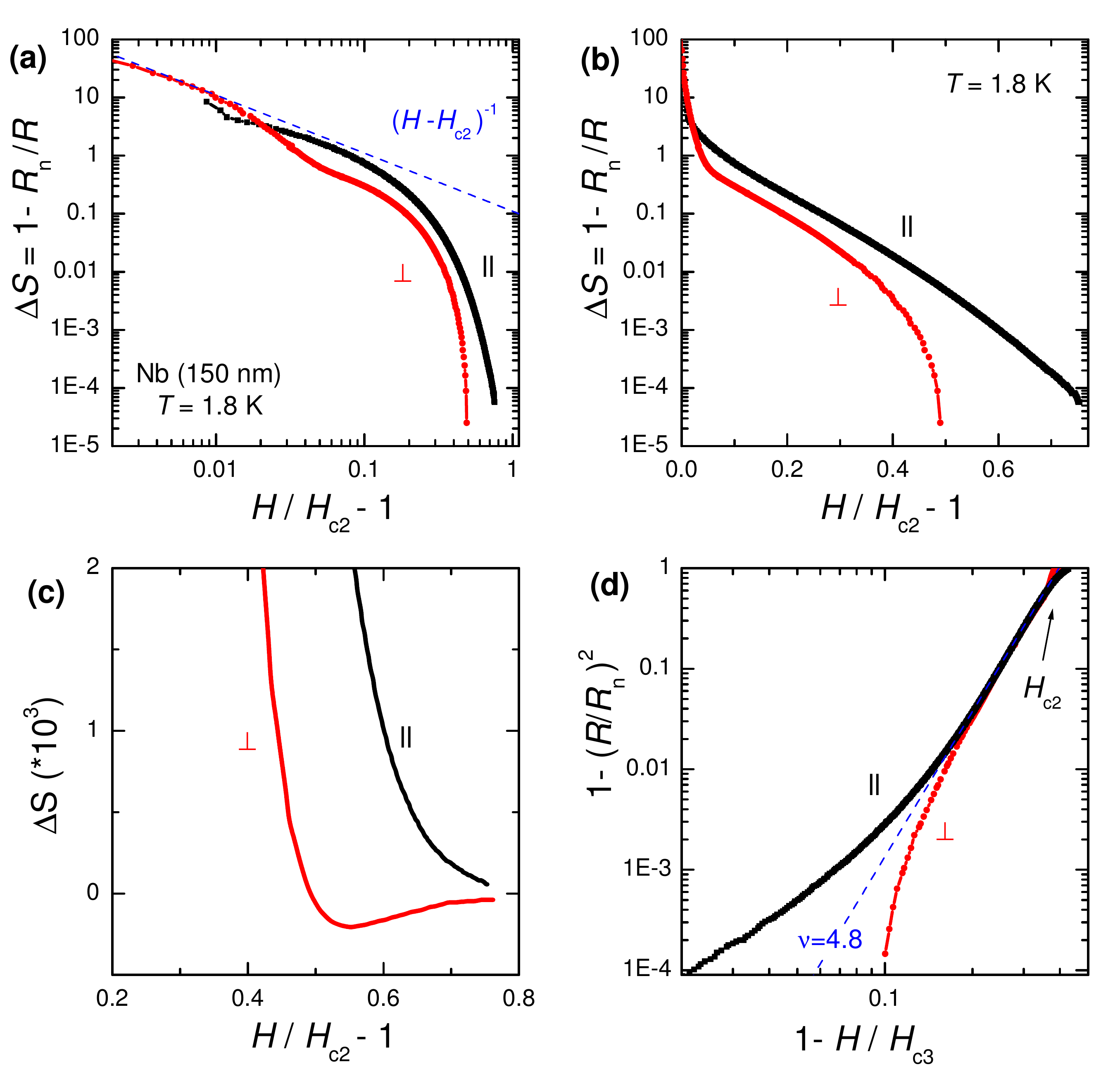}
    \caption{(Color online). (a) and (b) Excess conductance vs. magnetic field at $T=1.8$ K plotted in (a) a double-logarithmic and (b)
    a semi-logarithmic scale.
    Quasi-exponential decay $\Delta S(H)$ is seen. (c) The high-field part of excess
    conductance. It reveals fluctuation contribution, which is
    positive for parallel, but negative for perpendicular field
    orientation. (d) Analysis of power-law scaling expected for surface superconductivity
    according to Eq. (\ref{R_Q}). Dashed line represents a power-law with the exponent $\nu =4.8$.
    }
    \label{fig:fig5}
\end{figure*}

In Fig. \ref{fig:fig4} (c) we show $R(\Theta)$ curves for the SSC
model calculated from Eqs. (\ref{Hc3_Q}) and (\ref{R_Q}) for
$H=1.1H_{c2}$ at different bias. Calculations are made for
$H_{c3}^{\perp}=H_{c2}$ and $H_{c3}^{\parallel}=1.69 H_{c2}$ and
for $I_c(\Theta)=$const. For comparison we also show flux-flow
type dependence $R/R_n = H/H_{c3}$. It is seen that the cusp in
the SSC model is much sharper, primarily due to non-linearity of
the $I$-$V$. Overall behavior is similar to experimental data from
Fig. \ref{fig:fig4} (a), even though in experiment a very sharp
cusp at $\Theta=0$ survives up to much higher current. The
difference is due to an oversimplified assumption of
angular-independent $I_c(\Theta)=$const, used in calculations. In
reality $I_c(\Theta)$ has a sharp maximum at $\Theta=0$ because
the Lorenz force vanishes as $\sin(\Theta)$. It is possible to get
a better fit using a more realistic $I_c(\Theta)$, but we don't
want to go in to more complicated modelling because the main
purpose of calculations was to demonstrate how non-linearity of
$I$-$V$'s leads to a much sharper (compared to a simple 2D
flux-flow model) cusp in $R(\Theta)$.

\subsection{Analysis of fluctuation contribution}

Finally we discuss fluctuation contribution to the broadening of
resistive transition. In Fig. \ref{fig:fig5} (a) we show
normalized excess conductance $\Delta S= (1/R-1/R_n)R_n$ for the
data from Figs. \ref{fig:fig1} (b) and (c) in a double-logarithmic
scale. Such graph is usually used for analysis of fluctuation
contribution to conductivity. Dashed line shows $\Delta S \propto
(H-H_{c2})^{-1}$ dependence expected for 2D fluctuations
\cite{Varlamov_book}. It is seen that although there is a narrow
range of fields close to $H_{c2}$ with similar behavior, the
overall agreement is poor. In Fig. \ref{fig:fig5} (b) we replot
the same data in a semi-logarithmic scale. It is seen that $\Delta
S$ decays quasi-exponentially with increasing field at
approximately the same rate for both field orientations. A similar
exponentially decay versus both $T$ and $H$ has been reported for
high-$T_c$ cuprates \cite{MR,Katterwe_2014,Alloul2011,Luo2014}.
Here we demonstrate that it is generic also for conventional
superconductors. Such behavior is not expected for fluctuations
\cite{Varlamov_book,Finkelstein_2012,Glatz_2014} and we argue that
it is rather a signature of SSC.

It is possible to distinguish fluctuation contribution from
non-fluctuating SSC. SSC always leads to excess conductance, but
fluctuation contribution to magnetoresistance can be both positive
and negative \cite{Varlamov_book,Finkelstein_2012,Glatz_2014}. In
particular, at low $T$ and at field perpendicular to the film the
density-of-state contribution to fluctuations leads to excess
resistance, rather than excess conductance \cite{Glatz_2014}. We
clearly see such a contribution in our data. In Fig.
\ref{fig:fig5} (c) we show high-field part of excess conductance
in the linear scale. It is seen that in parallel field there is
always an excess conductance $\Delta S >0$, which rapidly
decreases upon approaching the surface critical field
$H_{c3}\simeq 1.7 H_{c2}^{\parallel}$, but never really vanishes.
The remaining tail is a signature of fluctuations that persist at
any field. For perpendicular field, $\Delta S$ at high fields
becomes {\it negative}, which is consistent with theoretical
expectations for fluctuation contribution at $T\ll T_c$
\cite{Glatz_2014}.

In Fig. \ref{fig:fig5} (d) we check the power-law scaling
suggested by Eq. (\ref{R_Q}) for SSC. It is seen that there is a
good scaling in a broad field range, although extraction of the
exponent $\nu$ is not very confident because it depends on the
chose of $H_{c3}$. The dashed line corresponds to $\nu=4.8$. Upon
approaching $H_{c3}$, deviations with opposite signs for parallel
and perpendicular field orientations appear, signalizing
fluctuation contributions. This indicates that SSC makes a
dominant contribution to excess conductivity at $H_{c2}<H\lesssim
0.8 H_{c3}$, while fluctuation contribution starts to become
significant only upon weakening of SSC at $H>0.8H_{c3}$ and takes
over completely at $H>H_{c3}$.

\section{Discussion}

Our results suggest that surface superconductivity is the primary
cause of broadening of superconducting transition in magnetic
field. As indicated in Figs. \ref{fig:fig1} (b) and (c),
$H=H_{c2}$ corresponds to the bottom of transition, consistent
with earlier studies \cite{Burger_1965}, and $H_{c3}$ to the top
of the resistive transition. Thus the full width of the transition
is dominated by SSC. Although SSC is well known for carefully
polished single crystals \cite{Park_2004,Das_2008}, it is usually
considered to be insignificant for disordered, rough or
inhomogeneous superconducting films because of its assumed
fragility and sensitivity to surface conditions
\cite{SaintJames,Casalbuoni_2005,Bellau_1966,Agterberg_1996}.
Therefore, observation of a very robust SSC in our strongly
disordered polycrystalline films is rather surprising, especially
for field perpendicular to the film. In perfectly uniform films
SSC should not occur at perpendicular field orientation
\cite{SaintJames,Abrikosov_1965}. Yet, SSC in perpendicular fields
has been directly visualized by scanning laser microscopy for
similar films \cite{Werner_2013} and also reported for some
layered supercondcutors \cite{LevyBertrand_2016} and sintered
polycrystalline MgB$_2$ samples \cite{Tsindlekht_2006}. Presumably
it is the polycrystallinity of our films that allows SSC at grain
boundaries even in perpendicular fields. Thus we conclude that
surface superconductivity is a robust phenomenon that should be
carefully considered in analysis of data close to superconducting
transition.

\begin{acknowledgements}

The work was supported by the Swedish Research Council Grant No.
621-2014-4314 and the Swedish Foundation for International
Cooperation in Research and Higher Education Grant No.
IG2013-5453.
\end{acknowledgements}


\end{document}